\documentclass[a4paper]{jpconf}
\usepackage{graphicx}
\begin{document}
\title{Mass and radius constraints for compact stars and the QCD phase diagram}

\author{David~B.~Blaschke}
\address{Institute for Theoretical Physics, 
	University of Wroclaw, 
	50-204 Wroclaw, Poland
\\
Bogoliubov Laboratory for Theoretical Physics, JINR Dubna
	141980 Dubna, Russia
}
\ead{blaschke@ift.uni.wroc.pl}

\author{Hovik~A.~Grigorian}
\address{Laboratory of Information Technologies, JINR Dubna
	141980 Dubna, Russia
\\
Department of Theoretical Physics,
	Yerevan State University, 
	0025 Yerevan, Armenia}
\ead{hovik.grigorian@gmail.com}

\author{David~E.~Alvarez-Castillo}
\address{Bogoliubov Laboratory for Theoretical Physics, JINR Dubna
	141980 Dubna, Russia
\\
Instituto de F\'isica, 
	Universidad Aut\'onoma de San Luis Potos\'i, 
	78290 San Luis Potos\'i, SLP, M\'exico} 
\ead{alvarez@theor.jinr.ru}

\author{Alexander~S.~Ayriyan}
\address{Laboratory of Information Technologies, JINR Dubna 
	141980 Dubna, Russia
}
\ead{ayriyan@jinr.ru}

\begin{abstract}
We suggest a new Bayesian analysis using disjunct M-R constraints for 
extracting probability measures for cold, dense matter equations of state. 
One of the key issues of such an analysis is the question of a
deconfinement transition in compact stars and whether it proceeds as a 
crossover or rather as a first order transition. 
The latter question is relevant for the possible existence of a critical 
endpoint in the QCD phase diagram under scrutiny in present and upcoming 
heavy-ion collision experiments.
\end{abstract}

\section{Introduction}
One of the most challenging problems of modern physics concerns the structure
of the phase diagram of quantum chromodynamics (QCD).
Experimental programs with heavy-ion collisions (HIC) at ultrarelativistic 
energies and large-scale simulations of lattice QCD at finite temperature are 
performed to identify the position and the character of the suspected 
transition from a gas of hadronic resonances to a quark-gluon plasma in these 
systems characterized by almost perfect symmetry between particles and 
antiparticles, i.e. vanishing baryon density.
It is nowadays established that there is a crossover transition with a 
pseudocritical temperature of 
$T_c=(154 \pm 9)$ MeV \cite{Bazavov:2011nk}
from lattice QCD and a chemical freeze-out temperature 
$T_{\rm fo}=(156 \pm 5)$ MeV \cite{Becattini:2012xb}
for all hadron species from a statistical model analysis of hadron 
production data at $\sqrt{s}=2.7$ TeV from the ALICE experiment at CERN LHC.

Both, HIC experiments and ab-initio lattice QCD cannot address, however,
the QCD phase diagram at low and vanishing temperatures where the QCD 
phase transition is eventually of first order so that a critical end point
(CEP) of first order transitions would result. 
The position of such a CEP in the phase diagram would be a landmark for the 
studies of strongly interacting matter under extreme conditions as
it could help to identify the universality class of QCD. 
However, the beam energy scan programs at RHIC (STAR experiment) or at 
CERN SPS (NA61 experiment) have not yet brought any evidence for the existence
of the CEP. 
Moreover, some theoretical studies argue that it could be absent at all because
of the persistence of repulsive vector meson mean fields in dense matter
\cite{Bratovic:2012qs}, see also the discussion in 
\cite{Contrera:2012wj}. 

In this situation, where effective model approaches give contradicting results,
and new experimental facilities like NICA at JINR Dubna and the CBM experiment 
at FAIR Darmstadt are not yet operative, a guidance for progress in the field 
could come from astrophysics of compact objects, namely from the precise mass
and radius measurement of pulsars  \cite{Miller:2013tca}.
The main question to be answered is
\cite{Blaschke:2013ana,Alvarez-Castillo:2013cxa}: 
Can there be a trace of a (strong)
first order phase transition in cold nuclear (neutron star) matter in 
mass-radius data from compact star observations?

In the present contribution, we outline a model study for this case which is 
based on Bayesian analysis (BA) methods and how they could guide future pulsar 
observational campaigns.

\section{EoS \& stars with a QCD phase transition}

For this study we follow the scheme suggested by Alford, Han and Prakash
\cite{Alford:2013aca} for the hybrid EoS with a first order phase transition,
\begin{equation}
\label{eq:hybrid}
 p(\epsilon)=p^{I}(\epsilon)~\Theta(\epsilon_{c}-\epsilon)
+p^{II}(\epsilon)~\Theta(\epsilon-\epsilon_c-\Delta \epsilon),
\end{equation}
where $p^{I}(\epsilon)$ is given by a pure hadronic EoS and $p^{II}(\epsilon)$ 
represents the high 
density nuclear matter introduced here as quark matter 
given in the bag-like form 
\begin{equation}
\label{eq:bag}
p^{II}(\epsilon)=c^{2}_{\rm QM}(\epsilon - \epsilon_0) 
=  c^{2}_{\rm QM}\epsilon - B,
\end{equation}
with $c^{2}_{\rm QM}$ as the squared speed of sound in quark matter, and the 
bag constant $B$, or the energy density offset $\epsilon_0$ being synonymous 
for parametrizing the latent heat $\Delta \epsilon$ of the phase transition, 
occuring at the critical pressure 
$p_c=p(\epsilon_c)=p^{I}(\epsilon_c)=p^{II}(\epsilon_c+\Delta \epsilon)$. 
It has been shown by Haensel et al.~\cite{Zdunik:2012dj} 
that Eq.~(\ref{eq:bag})
describes pretty well the superconducting NJL model derived 
in~\cite{Blaschke:2005uj} and applied for hybrid stars with an extension by a
repulsive vector meanfield first in \cite{Klahn:2006iw}, recently revisited 
and systematically sampled in \cite{Klahn:2013kga}.

For the hadronic EoS we take the well known model of APR~\cite{Akmal:1998cf} 
that is in agreement with experimental data at densities about nuclear 
saturation. 
For this hadronic branch (I) all the relevant thermodynamical variables, 
energy density $\epsilon$, pressure $p$, baryon density $n$ and chemical 
potential $\mu$ are well defined and taken as input for determination of 
the hybrid (hadronic + quark matter) EoS. 

\begin{figure}[ht!]
\begin{center}
\includegraphics[width=0.48\textwidth]{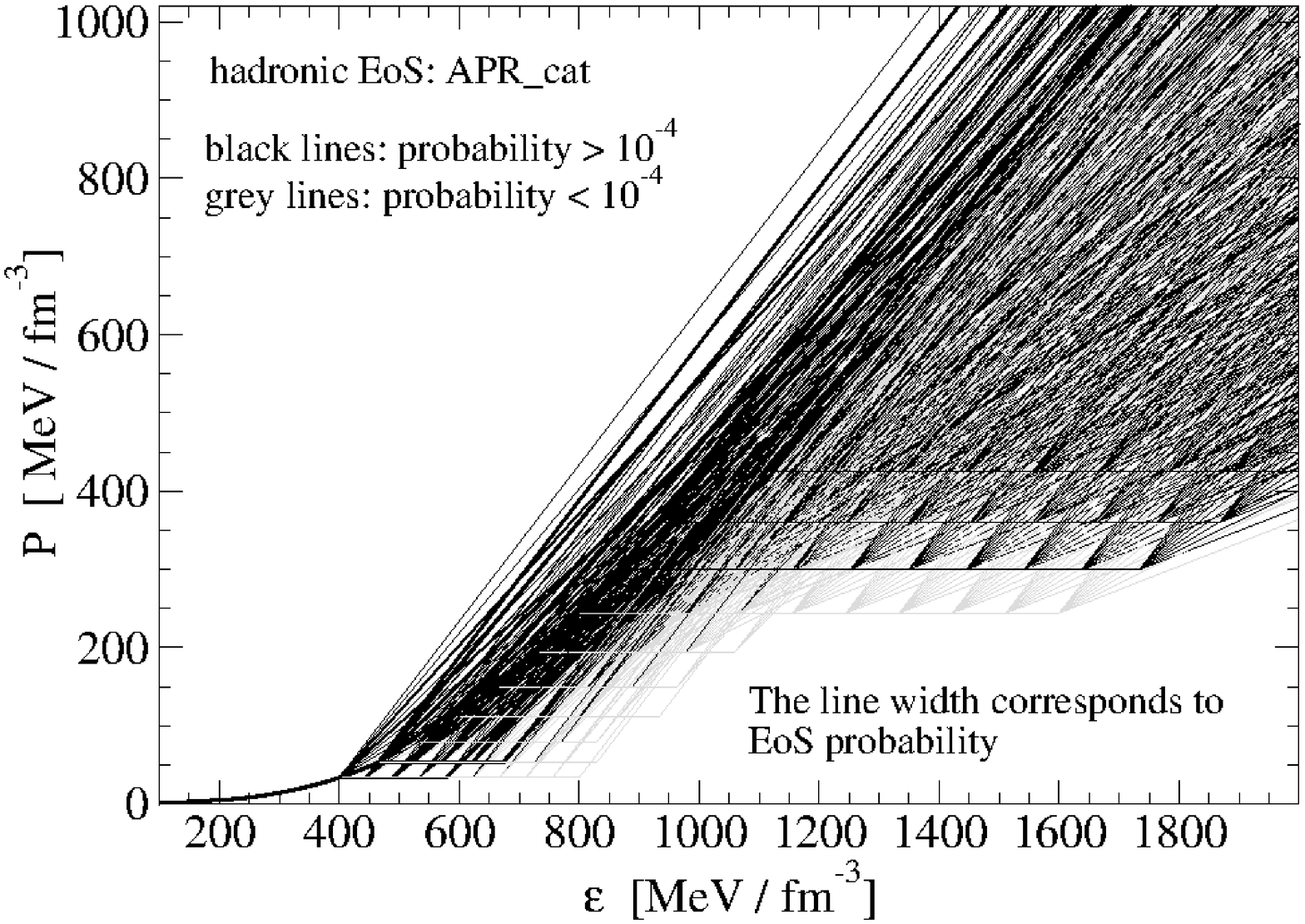}
\hfill
\includegraphics[width=0.45\textwidth]{PvsEps_APR_cat_10x10x10_lin.eps}
\\
\includegraphics[width=0.48\textwidth]{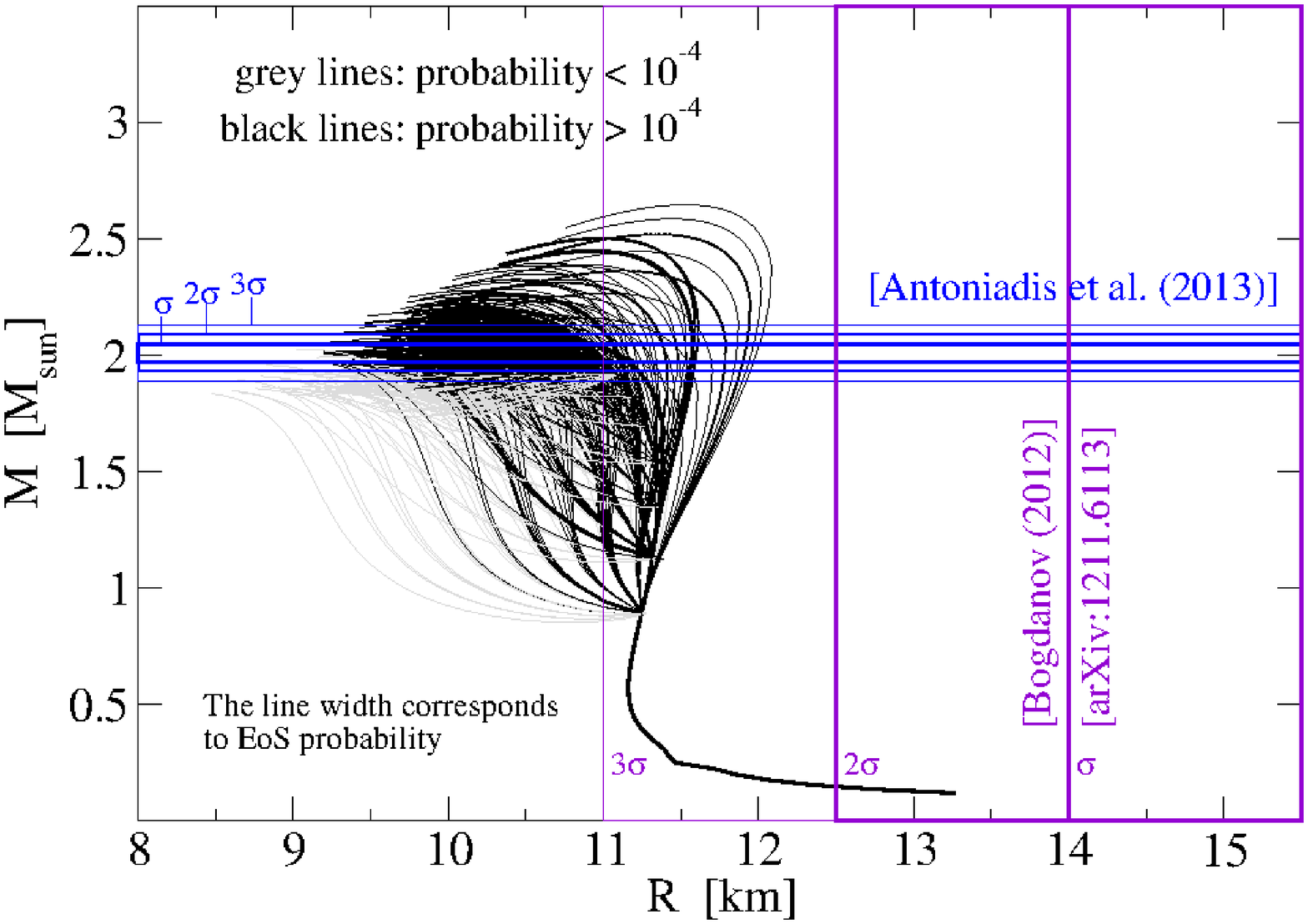}
\hfill
\includegraphics[width=0.45\textwidth]{MvsR_APR_cat_10x10x10.eps}
\\
\includegraphics[width=0.45\textwidth]{MvsM_B_APR_cat_1000.eps}
\hfill
\includegraphics[width=0.45\textwidth]{MvsM_B_APR_cat_10x10x10.eps}
\end{center}
\caption{Hybrid EoS scheme for sets of three parameters 
$\left(\epsilon,\gamma,{c}_{s}^{2}\right)$ before (left panels) and 
after (right panels) the Bayesian analysis.}
\label{AHP_Scheme}
\end{figure}

The free parametes of the model are the transition density $\epsilon_c$, the 
energy density jump $\Delta \epsilon \equiv \gamma \epsilon_c$ and 
$c^{2}_{\rm QM}$. 
For the present study we will use all hybrid EoS of the given type which 
are obtained when varying these three parameters within the limits:
$400<\epsilon_c$[MeV/fm$^3$]$<1000$, $0< \gamma < 1.0$ and 
$0.3<c^{2}_{\rm QM}<1.0$. 
The resulting EoS are shown in the upper left panel of Fig.~\ref{AHP_Scheme} 
and demonstrate which part of the pressure versus energy density plane is 
covered by our three-dimensional parameter sampling.
For the present study we use a set of 1000 EoS corresponding to a coverage of 
the parameter space by 10 values in each dimension. 

Using the above set of hybrid EoS one calculates the 
corresponding set of neutron star sequences by solving the 
Tolman-\-Oppenheimer-\-Volkoff (TOV) equations \cite{Glendenning:1997wn}.
Of particular interest for the comparison with observational data are the
gravitational mass vs. radius ($M-R$) and 
gravitational mass vs. baryon mass ($M-M_B$) diagrams, shown in 
Fig.~\ref{AHP_Scheme} in the middle and lower left panels, respectively.
One can read off the following correlations with the hybrid EoS parameters:
(i) the higher the critical energy density ($\epsilon_c$) the higher the onset 
mass for hybrid star configurations,
(ii) the larger the jump in energy density at the transition ($\gamma$) the
stronger the effect of compactification of the hybrid star configuration, 
which is reflected also in a larger gravitational binding energy (mass defect)
for the rightmost curves in the $M-M_B$ diagram (light color). 
Increasing $\gamma$ eventually leads to an instability, indicated by an 
increasing mass with increasing radius, and  
(iii) the increase of the maximum mass for a star sequence which results from
increasing the stiffness of quark matter, i.e. the speed of sound 
$c^{2}_{\rm QM}$. 

In the next step, we would like to apply observational constraints to the 
obtained 1000 compact star sequences and filter the most likely parameter 
sets by applying Bayesian methods.

\section{Observational constraints and Bayesian analysis}

We want to compare the theoretical results with suitable constraints from 
neutron star observations in order to conclude for the likeliness of the 
underlying EoS model. 
A pioneering study of this kind has been 
performed recently in \cite{Steiner:2010fz}. 
However, as 
there are suspicions that the analysis of the burst sources used in 
\cite{Steiner:2010fz} may 
have a systematic bias towards smaller radii 
(cf. \cite{Miller:2010,Trumper:2011}),
we use here instead three statistically independent constraints, none of them 
related to bursting sources:
\begin{itemize}  
\item a maximum mass constraint from PSR J0348+0432~\cite{Antoniadis:2013pzd}, 
\item a radius constraint from the nearest millisecond pulsar 
PSR J0437-4715~\cite{Bogdanov:2012md}, which both are shown in the 
middle panel of Fig.~\ref{AHP_Scheme}, and 
\item a constraint on the gravitational binding energy from the neutron star B 
in the binary system J0737-3039 (B)~\cite{Kitaura:2006}, see the bottom 
panel of Fig.~\ref{AHP_Scheme}. The $M_B$ from this full hydro simulation is
1\% smaller than the one by \cite{Podsiadlowski:2005ig} where mass loss was 
neglected.
\end{itemize}
The above constraints are shown with their respective $1\sigma-$, $2\sigma-$,
and $3\sigma-$ confidence regions in the lower panels of Fig.~\ref{AHP_Scheme}.
For the BA, we have to sample the above defined parameter space
and to that end we introduce a vector of the parameter values
\begin{equation}
\label{pi_vec}
\pi_i=\overrightarrow{\pi}\left(\epsilon_c(k),\gamma(l),c_{\rm QM}^2(m)\right),
\end{equation}
where 
$i = 0\dots N-1$ (here $N = N_1\times N_2\times N_3$) as 
$i = N_1\times N_2\times k + N_2\times l + m$ and 
$k = 0\dots N_1-1$, $l = 0\dots N_2-1$, $m = 0\dots N_3-1$. 
Here $N_1$, $N_2$ and $N_3$ denote the number of parameters 
$\epsilon_c$, $\gamma$ and $c^2_{\rm QM}$, respectively.
The goal is to find the set $\pi_i$ corresponding to an EoS and thus a 
sequence of configurations which contains the most probable 
one based  on the given constraints using BA.
For initializing the BA we propose that {\it a priori} each vector of 
parameters $\pi_i$ has a probability equal to unity: 
$P\left(\pi_i\right) = 1$ for $\forall i$.
Then one proceeds as follows.

\subsection{Mass constraint}
We propose that the error of mass measurement is normal distributed 
$\mathcal{N}(\mu_A,\sigma_A^2)$, where $\mu_A = 2.01~\mathrm{M_{\odot}}$ and 
$\sigma_A = 0.04~\mathrm{M_{\odot}}$, according to the mass measurements 
for the massive pulsar PSR~J0348+0432~\cite{Antoniadis:2013pzd}. 
Using this assumption we can calculate the conditional probability of the 
event $E_{A}$ that the mass of a neutron star corresponds to this measurement
\begin{equation}
\label{p_anton}
P\left(E_{A}\left|\pi_i\right.\right) = \Phi(M_i, \mu_A, \sigma_A),
\end{equation}
where $M_i$ - maximal mass constructed by $\pi_i$ and $\Phi(x, \mu, \sigma)$ 
is the cumulative distribution function for the normal distribution.

\subsection{Radius constraint}
From an analysis of the timing of the nearest millisecond pulsar PSR~J0437-4715
Bogdanov~\cite{Bogdanov:2012md} extracts a radius of 
$\mu_B = 15.5~\mathrm{km}$ 
at a mass of $1.7~M_\odot$ with a variance of $\sigma_B = 1.5~\mathrm{km}$.
We will consider this value mass independent, neglecting the mild variation 
given in Ref.~\cite{Bogdanov:2012md} since it is inessential for the present 
study.
Now it is possible to calculate conditional probability of the event $E_{B}$ 
that the radius of a neutron star corresponds to the given measurement
\begin{equation}
\label{p_bogdan}
P\left(E_{B}\left|\pi_i\right.\right) = \Phi(R_i, \mu_B, \sigma_B)~.
\end{equation}

\subsection{Gravitational binding ($M-M_B$) constraint}
This constraint 
gives a region in the $M-M_B$ plane. 
For our analysis we use the mean values $\mu = 1.249~M_\odot$, 
$\mu_B = 1.360~M_\odot$ and 
the standard deviations $\sigma_{M} = 0.001~M_\odot$ and 
$\sigma_{M_B} = 0.002~M_\odot$ 
which are given in \cite{Kitaura:2006}. 

We need to estimate the probability for the closeness of a theoretical point 
$M_i = \left({M}_i, {M_B}_i\right)$ to the observed point 
$\mu = \left(\mu, \mu_B\right)$. 
The required probability can be calculated using the following formula
\begin{equation}
\label{p_kitaura}
P\left(E_{K}\left|\pi_i\right.\right) = 
\left[ \Phi\left(\xi\right) 
- \Phi\left(-\xi\right) \right]\cdot\left[ \Phi\left(\xi_B\right) 
- \Phi\left(-\xi_B\right) \right],
\end{equation}
where $\Phi\left(x\right) = \Phi\left(x, 0, 1\right)$, 
$\xi = {\sigma_{M}}/{d_{M}}$ and 
$\xi_B = {\sigma_{M_B}}/{d_{M_B}}$, with $d_{M}$ and $d_{M_B}$ being the 
absolute values of components of the vector 
$\mathrm{\textbf{d}_i} = \mathrm{\bf\mu} - \mathrm{\textbf{M}}_i$, where 
$\mathrm{\bf\mu} = \left(\mu, \mu_B\right)^T$ is given in 
\cite{Kitaura:2006} 
and $\mathrm{\textbf{M}}_i = \left({M}_i, {M_B}_i\right)^T$ is the solution 
of the TOV equations using the $i^{\mathrm{th}}$ vector of EoS parameters 
$\pi_i$.
Note that formula (\ref{p_kitaura}) does not correspond to the multivariate 
normal distribution.

\subsection{Calculation of {\it a posteriori} probabilities}
Note, that these measurements are independent of each other. 
This means that we can calculate the complete conditional probability of 
an event described by one of the objects in the star sequence addressed by 
$\pi_i$ corresponds to a product of 
the probabilities for all three constraining measurements
\begin{equation}
\label{p_event}
P\left(E\left|\pi_i\right.\right) = 
P\left(E_{A}\left|\pi_i\right.\right) 
\times P\left(E_{B}\left|\pi_i\right.\right) 
\times P\left(E_{K}\left|\pi_i\right.\right).
\end{equation}

\section{Results and conclusions}

As a result of the BA for the three-parameter EoS (\ref{eq:hybrid}) each of 
the $10 \times 10 \times 10 = 1000$ points in the three-dimensional parameter 
space has been assigned a probability value according to (\ref{p_event}).
We have selected the six most likely parameter sets and show the corresponding
equations of state in the upper right panel of Fig.~\ref{AHP_Scheme}. 
The corresponding $M-R$ and $M-M_B$ plots are given in the lower right panels
of Fig.~\ref{AHP_Scheme}, together with the corresponding observational 
constraints.

\begin{figure}[ht!]
\includegraphics[width=0.9\textwidth]{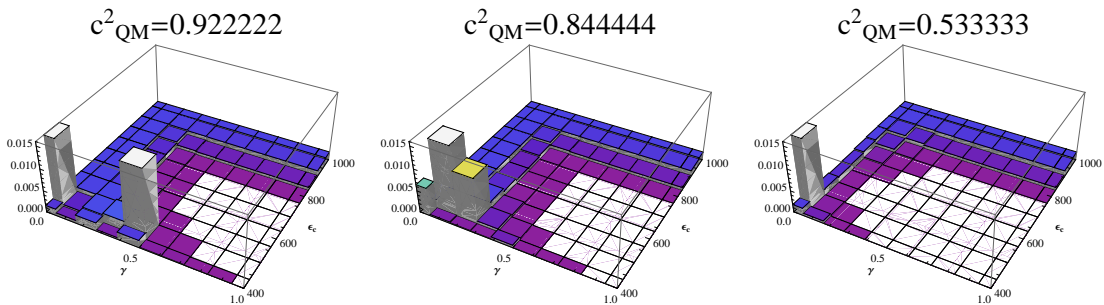}
\caption{LEGO plots for Bayesian analysis in the plane of two parameters 
$\left(\epsilon_c,\gamma\right)$ for three fixed values of the third parameter
${c}_{s}^{2}$.}
\label{Bayesian}
\end{figure}

Alternativey, one can present the probabilities as histograms in the parameter 
space. Here, we are restricted to a two-dimensional subset (LEGO plots).
Therefore we choose the three classes of speed of sound values for which we 
found the maximum probabilities and depict for each of them the LEGO plot of
probabilities in the subspace of critical energy density ($\epsilon_c$) and 
strength of the first-order transiition ($\gamma$) in Fig.~\ref{Bayesian}.
It is interesting to note that for the stiffest high-density EoS 
($c^2_{\rm QM}=0.922$), we obtain a bimodality of the probability distribution.
A crossover behaviour ($\gamma=0$) has the same probability as a strong 
first order transition ($\gamma=0.5$).

We conclude that the Bayesian Analysis we have presented here presents an 
alternative to the previously developed one in Ref.~\cite{Steiner:2010fz}.
It is based on three statistically independent constraints which are extremely
selective and result in maximum probabilities of about 1\% for the most likely
EoS. 
Thus this method will have a sufficient selective power when applied to a 
broader class of EoS.
This concerns in particular EoS being stiffer (with larger radii) on the 
hadronic side and with microscopically founded EoS on the high density 
(quark matter) side. 
To test the interesting possibility that a strong first order phase transition 
might occur in massive neutron stars of $2~M_\odot$ 
\cite{Antoniadis:2013pzd,Demorest:2010bx} and lead to the appearance of a 
``third family'' (mass twins) of hybrid stars at this high mass, one should 
perform radius measurements for these massive neutron stars.
If it might turn out that the corresponding radii might be significantly 
different, e.g., by $1-2$ km with a standard deviation of about 500 m, then 
one would be able to disselect EoS without a phase transition, given the 
very narrow range of uncertainty in the mass of these objects.
This possibility offers bright prospects for future observational campaigns and
bears the chance to ``prove'' the existence of a critical point 
\cite{Blaschke:2013ana,Alvarez-Castillo:2013cxa} 
in the QCD phase diagram from astrophysical observations!
 
\section*{Acknowledgements}
D.B., H.G. and D.E.A.-C. acknowledge the stimulating and inspiring atmosphere 
at the workshop in Yerevan, as well as support 
by the Volkswagen Foundation.
The authors are grateful for clarifying discussions with many colleagues about 
this topic, in particular with T. Fischer, T. Kl\"ahn, R. Lastowiecki, 
J. Lattimer, M.C. Miller, J. Tr\"umper, F. Weber.
The work was supported by the 
NCN ``Maestro'' programme under contract number UMO-2011/02/A/ST2/00306.
A.S.A.,  D.E.A-C. and H.G. are grateful for support by the Bogoliubov-Infeld 
programme.
A.A. thanks to JINR grant No. 14-602-01 and H.G. acknowledges support by the Volkswagen Foundation under grant No. 85 182.
D.E.A-C. is grateful for support from funds of 
the collaboration between South Africa and JINR Dubna for
his participation at the School on Bayesian Analysis in Physics and Astronomy
at the University of Stellenbosch and the SKA Conference at the STI$\alpha$S.

\end{document}